\input{aipcheck}

\documentclass[
    ,final            
  ]
  {aipproc}

\layoutstyle{6x9}

\newcommand{\jp}{J/\psi}
\newcommand{\bzb}{{\bar{B}}^0}
\newcommand{\bz}{B^0}
\newcommand{\ks}{K_S}
\newcommand{\kl}{K_L}
\newcommand{\ra}{\rightarrow}
\newcommand{\Eb}{E_{\rm beam}}
\newcommand{\EB}{E_{B^0}}

\begin{document}

\title{CP Violation and the Future of Flavor Physics}


\author{Christian Kiesling}{
  address={Max-Planck-Institute for Physics, Munich, Germany}
}

\begin{abstract}
 With the nearing completion of the first-generation experiments at
 asymmetric $e^+ e^-$ colliders running at the $\Upsilon(4S)$ resonance
 (``B-Factories'') a new era of high luminosity machines is at the
 horizon. We report here on the plans at KEK in Japan to upgrade the
 KEKB machine (``SuperKEKB'') with the goal of achieving an
 instantaneous luminosity exceeding $8 \times 10^{35}$ cm$^{-2}$
 s$^{-1}$, which is almost two orders of
 magnitude higher than KEKB. Together with the machine, the Belle
 detector will be  
 upgraded as well (``Belle-II''), with significant improvements to
 increase its background tolerance as well as improving its physics
 performance. The new generation of experiments is scheduled to take
 first data in the year 2013.

\end{abstract}

\maketitle


\section{Introduction}

The Belle Collaboration, together with BaBar, has made essential
contributions to establish the theory of Kobayashi and Maskawa, who
explain all known CP violation phenomena within the framework of the
Standard Model (SM) by a single irreducible phase appearing in the
quark mixing matrix. For their outstanding achievement Kobayashi and
Maskawa were awarded the Nobel Prize of 2008. Although the SM has
been extremely  
successful in describing virtually all data, most importantly the CP
violation phenomena of the $K$- and the $B$-systems (for a
comprehensive overview see, e.g.~\cite{buras}), there are a
number of arguments why the SM cannot be the regarded as a complete
theory. In fact, there is clear evidence for physics beyond the SM, as
suggested by the non-vanishing mass of the neutrinos, the
matter-antimatter asymmetry observed in the universe, and the apparent
necessity for dark matter. Most likely, the effects mentioned have to
do with CP violation of a yet undiscovered source. The ``New Physics'' (NP)
generating these sources is expected to appear at large, so far
unreached (multi-TeV) energy scales. While the discovery and
exploration of New  
Physics is the central motivation for the LHC program, flavor physics is
expected to play a key role in unraveling possible NP at this scale
and to solve the puzzle of CP violation.

\begin{figure}
  \includegraphics[height=.17\textheight]{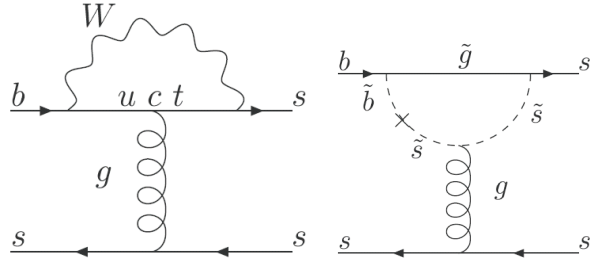}
  \caption{\label{fig:penguin}Example of a SM process at the quantum
  loop level (``penguin diagram'', left) with ``New Physics'' contributing
  (right) to $B$-meson decay amplitudes.}
\end{figure}

Colliding $e^+$ and $e^-$ beams with different (``asymmetric'') energies 
to produce the $\Upsilon(4S)$ resonance, with
just enough energy for the creation of a pair of $B$-mesons (or background
with other quark flavors), is an alternative approach to the
high-energy frontier experiments at the LHC. With high luminosity, and
consequently large statistics, as achieved in the next generation of
flavor factories (``Super Flavor Factory'',
SFF)~\cite{superkekb,superb}, very large energy 
scales can be reached, when quantum loop corrections to the SM are
considered (see fig.~\ref{fig:penguin}). Depending on the flavor
changing couplings of the NP particle spectrum, the sensitivity to large
mass scales in a SFF may be from many hundred GeV
up to tens of TeV. In this respect a SFF is truely complementary to
the LHC. A recent review of the physics potential of a future high
luminosity B factory can be found, e.g., in~\cite{newphys}. One should note
here that the discovery potential of a future SFF is indeed
extraordinary and might reach even beyond the LHC.

There are several distinct approaches to look for NP at
SFF's: Many of them concentrate on the precise measurement of the
angles and sides of the unitarity triangle for $B$-meson couplings,
derived from the Cabibbo-Kobayashi-Maskawa (CKM) matrix which connects
the mass eigenstates to the flavor eigenstates of the 
down-type quarks. Within the SM the CKM matrix is unitary and their
elements describe the coupling strengths of the flavor changing
currents at the quark level, such as $b \ra c$ or $b \ra u$. The
unitarity of the CKM 
matrix gives rise to a total of six 
so-called unitarity triangles, one of them involving the $b$-quark
couplings (``$B$-triangle''). All triangles have the same area, but the
$B$-triangle has all three sides of the same order, corresponding to
large angles and thus giving rise to large CP violating effects. Within
the SM the 
$B$-triangle is highly over-constrained (5 observables for only 2
independent quantities), so a precise measurement of all the three
angles and the two sides is a crucial check of the validity of the SM:
If the $B$-triangle ``does not close'', New Physics must be the
reason. One should mention that the all present measurements of the
CKM unitarity are in agreement with the SM, although a few ``tensions''
have become noticeable (for details see, e.g. ~\cite{CKMfitter}).
   
Another way of gaining sensitivity at the SFF to NP is the study of
rare decays of $B$ mesons and $\tau$ leptons. Some examples should
illustrate this point, such as $B \rightarrow X_d \nu \bar{\nu}$ or
$\tau \rightarrow \mu \gamma$. These decays involve (several) neutral
particles in the final state and can therefore only be measured at
SFF's. Such decays are highly suppressed (in the $B$ case) or even
completely forbidden (the $\tau$ case). With a SFF, branching fractions
down to several $10^{-9}$ can be probed. More details on rare
decays and their potential to search for New Physics can be found in
the proposals for the SFFs~\cite{superkekb,superb}.

One of the flagship measurements at the SFF is the precise
determination of the time-dependent CP violating asymmetries. These
asymmetries are measured by observing the decay rate $\Gamma$, as
function of time, of a $\bz$ meson decaying into a specific CP eigenstate
$f_{CP}$, as compared to the same final state coming from the
$\bzb$. The CP violating time-dependent asymmetry is defined as:
$$
{\cal A}(f_{CP},\Delta t) =\frac{\Gamma(\bzb \ra f_{CP};\Delta
t)-\Gamma(\bz \ra f_{CP};\Delta t)}{\Gamma(\bzb \ra f_{CP};\Delta
t)+\Gamma(\bz \ra f_{CP};\Delta t)}, 
$$
where $\Gamma$ is the decay rate of the $\bz(\bzb)$ into the CP
eigenstate $f_{CP}$ (``CP side'') within some time interval $\Delta
t$, to be explained below. In order to determine which of the
two flavors 
($\bz$ or $\bzb$) has decayed into the common final state $f_{CP}$,
the other $B$-decay (``tag side'') is analyzed for a specific flavor,
using, e.g., semi-leptonic decays ($\bz \ra X l^- \bar{\nu}, \bzb \ra
X l^+ \nu$). The 
charge of the lepton uniquely identifies (``tags'') the flavor of the
$B$ meson, where a positive (negative) lepton signals a
$\bzb$($\bz$). Since the two $B$ mesons are produced in an entangled
state by virtue of the quantum numbers of the $\Upsilon (4S)$, the
tag side uniquely\footnote{strictly speaking, the quantum entanglement
is broken when the first of the $B$-mesons decays. From then on the
other $B$-meson is freely oscillating between $\bz$ and $\bzb$, but with
a known time dependence.} determines the flavor of the $B$-meson that 
decayed into the CP 
eigenstate. The time difference $\Delta t$ is given by the difference
in decay times between the tag side and the CP side. Note that $\Delta
t$ can be positive or negative.

\begin{figure}
  \includegraphics[height=.3\textheight]{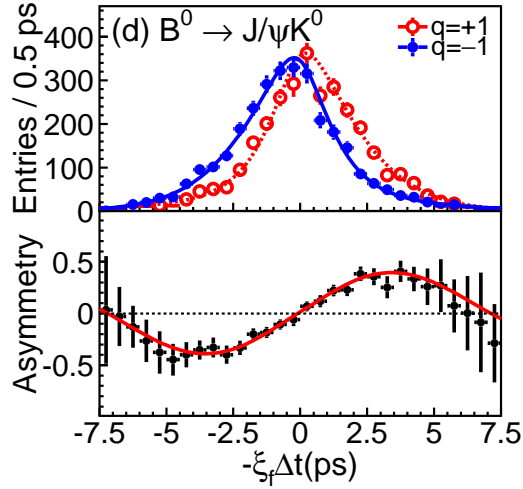}
  \caption{Recent measurements of the time-dependent CP asymmetry
  using the final states $\jp \ks$ and $\jp \kl$ (see~\cite{jpsiks}).}

\label{fig:jpsiksacp}
\end{figure}

Depending on the CP eigenstate chosen, any of the three angles $\phi_1
(\beta), \phi_2(\alpha)$ or $\phi_3 (\gamma)$ of the $B$-triangle can
be measured. Chosing, 
e.g., the final state $\jp K^0$, the angle $\phi_1$ (or $\beta$) is
determined. A recent measurement~\cite{jpsiks} of the time-dependent CP
asymmetry for the $\jp K^0$ channel is shown in
fig.~\ref{fig:jpsiksacp}. Here, the asymmetry is given for both odd
($\jp \ks$) and even ($\jp \kl$) CP final states. Since the $\Delta t$ 
distributions for the CP eigenstates are different, CP violation 
is established. Recent summaries on the measurements of the angles
$\phi_1, \phi_2, \phi_3$ ($\alpha, \beta, \gamma$) have been presented at
this conference~\cite{angle_summ}.     

\section{Machine Upgrade: SuperKEKB}

While no significant deviations from the SM predictions have
been observed so far, there are some tantalizing hints
for possible New Physics in $B$ decays (see,
e.g.,~\cite{hints}). Clarification can 
only come with a new generation of SFF's, which should aim at
integrated luminosities in excess of 50 /ab (the present world record
KEKB accelerator is about to accumulate 1 /ab). Such large integrated
luminosities require an equally large increase of the instantaneous
luminosity ${\cal L}$ which, in its simplified form, is given by
$$
{\cal L}=\frac{N_1N_2f}{4\pi\sigma_x\sigma_y}.
$$
Here, $N_i$ are the numbers of particles in each of the two colliding
bunches, $f$ is the bunch collision frequency, and $\sigma_{x,y}$ are
the transverse dimensions of the colliding bunches. 

\begin{figure}
  \includegraphics[height=.35\textheight]{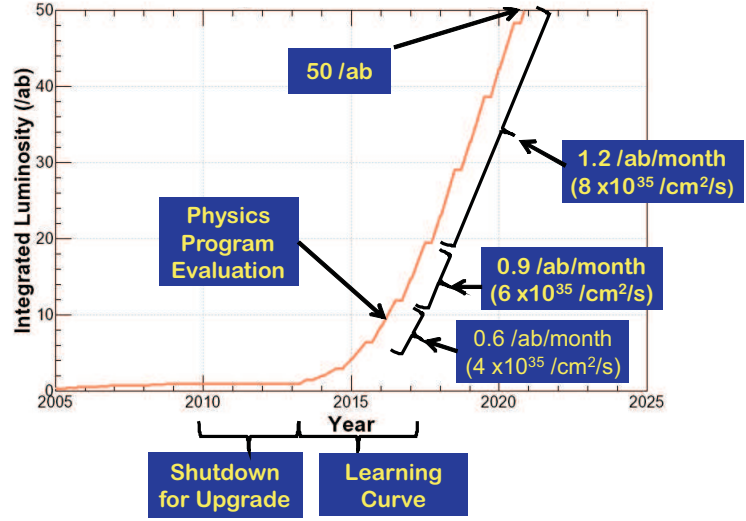}
  \caption{Expected development of the luminosity for the SuperKEKB
  machine.} 
\label{fig:lumidev}
\end{figure}

At KEK, an extremely strong
accelerator research program is focussing on an asymmetric $e^+e^-$
collider with instantaneous luminosities in excess of 8 $\times
10^{35}$ cm$^{-2}$ s$^{-1}$ (which is about 40 times the present
world record luminosity of 2.11 $\times 10^{34}$ cm$^{-2}$ s$^{-1}$,
reached in May 2009 with KEKB). The new machine, called ``SuperKEKB'',
is an upgrade of the present KEKB machine and should start producing
luminosity by the year 2013. According to the current plan the
KEKB accelerator should stop running by the end of 2009, so that the
construction work for SuperKEKB can start in 2010. The expected
luminosity development of the SuperKEKB machine is shown in
fig.~\ref{fig:lumidev}. 

Two options for the SuperKEKB machine have been discussed: The
``high-current'' (HC) option and the ``nano-beam'' (NB)
option (see Table~\ref{tab:machine}). Initially, the HC option was
favored. It was characterized by 
a mild decrease of the beta function in the low energy ring (LER), and
a dramatic increase of the beam current in the LER and the high energy
ring (HER). In addition, an ingenious scheme was developed to counteract
the luminosity loss due to the finite angle under which the two
colliding bunches cross each other, i.e. the crab crossing scheme. In
this scheme special crab cavities before and after the interaction
region rotate the bunches, so that they collide head-on, instead of at
an angle. The crab crossing scheme was verified with the running KEKB
accelerator, increasing the instantaneous luminosity by about 20
percent to a new world record. However, it became apparent that a new
effect would make the HC option difficult: For the crab crossing to
work efficiently the bunch length must be smaller than the $\beta$
function at the interaction point (IP), i.e. $\sigma_z
< \beta^*_y$. This condition, imposed to avoid the so-called hour-glass
effect, creates no problems at low beam currents. However, at 
the large beam currents of the HC option (see 
Table~\ref{tab:machine}) coherent synchrotron radiation becomes an
issue which has the effect of lengthening the bunch, thus running into
the hour-glass effect. The unavoidable bunch lengthening leads to a
decrease of the obtainable maximum luminosity to slightly over 50 
$\times 10^{34}$. Failing the goal of ${\cal L} \ge 80 \times
10^{34}$, this option is disfavored now. 

Following the ideas of the final focus system envisaged for a
future linear collider, transferred to a circular machine as
laid down in~\cite{superb}, the NB option has become the
baseline~\cite{machine_par}. Here, the key parameters are a factor of
two increase in the  
beam currents with respect to the present KEKB, but strongly reduced
$\beta$ 
functions at IP (factor of about 20 smaller than in the HC option). In
addition, low emittance beams are necessary to achieve the desired
small (less than 100 nano meters) transverse bunch sizes. Such a beam
size has been achieved at the ATF damping ring facility at KEK. While
a low emittance electron beam can be prepared by virtue of a carefully
designed injection system, the transverse phase-space of the positrons
needs to be cooled in a special new damping ring.           

\begin{table}
\begin{tabular}{l|c|c|c|c|}
\hline
  & \tablehead{1}{c}{b}{KEKB  \\ Design}
  & \tablehead{1}{c}{b}{KEKB achieved \\ (): with crab}
  & \tablehead{1}{c}{b}{SuperKEKB \\ High-Current Option}
  & \tablehead{1}{c}{b}{SuperKEKB \\ Nano-Beam Option} \\
\hline
$\beta^*_y$ (mm) (LER/HER) & 10 / 10 & $\begin{array}{c} 6.5 / 5.9 \\
(5.9 / 5.9) \end{array}$ & 3 / 6  & 0.26 / 0.26 \\

\hline
$\epsilon_x$ (nm) & 18 / 18 & 18 / 24 & 24 / 18  & 2.8 / 2.0 \\
\hline
$\sigma_y$ ($\mu$m) & 1.9 & 1.9 (0.94) & 0.85 / 0.73  &  0.073 / 0.097 \\
\hline
$\xi_y$ & 0.052 & $\begin{array}{c}0.108 / 0.057 \\ (0.129 /
0.090) \end{array}$ & 0.3 / 0.51  & 0.079 / 0.079 \\

\hline
$\sigma_z$ ($\mu$m) & 4 & ~7 & 5 (LER) / 3 (HER)  & 5 \\
\hline
$I_{\rm beam}$(A) & 2.6/1.1& $\begin{array}{c}1.66 / 1.34 \\ (1.64 /
1.19) \end{array}$ & 9.1 / 4.1 & 3.84 / 2.21 \\

\hline
$N_{\rm bunches}$ & 5000 & 1388 (1585) & 5000 & 2252 \\
\hline
$\begin{array}{l}{\rm Luminosity} \\ (10^{34} {\rm cm}^{-2} {\rm
s}^{-1}) \end{array} $& 1 & 1.76 (2.11) & 53 & 80 \\

\hline
\end{tabular}
\caption{Comparison of parameters for the two high luminosity options
of the SuperKEKB machine, compared to the presently running KEKB
factory. The high current option for SuperKEKB has been discarded
recently (see text) and the nano-beam option is now the baseline.} 

\label{tab:machine}
\end{table}

Table~\ref{tab:machine} gives an overview of the machine parameters
for the presently running KEKB accelerator, and for both the HC and NB
options (status of July 2009). Note that the presently achieved
luminosity of KEKB is a factor of two larger than  
the original design value. Part of
it is due to the new crab crossing scheme, which is being successfully
tested with the running machine. It should be noted, however, that the
luminosity increase predicted in simulation for the crab crossing
scheme has not yet 
been reached in the real machine. The discrepancy is attributed to
so-called ``machine errors'' which may result from a finite precision in
aligning the focussing quadrupoles along the ring. Small (unknown)
deviations from the ideal position can lead to a coupling of the
horizontal and vertical betatron oscillations and therefore to an
increase in the vertical beam size at the IP, reducing the
luminosity. During the last winter shutdown some errors of this kind
have been diagnosed and corrected for by installing a set of so-called
``skew sextupoles'' around the ring. As a consequence, the luminosity
could be immediately increased by more than 15 percent, leading to
new world records for the instantaneous luminosity (see
table~\ref{tab:machine}). 

Preliminary parameters of both collision schemes for the SuperKEKB
machine are given in the table~\cite{machine_par}. Intense
optimization studies for 
the new machine are going on, concentrating now mainly on the NB
option. One should note, however, that the parameters given in the
table have not yet been consolidated. In addition, the so-called crab 
waist scheme, as proposed in~\cite{superb}, has not yet been included
in the present design. While the crab waist scheme is expected to
improve the instantaneous luminosity only by a few percent, it is very
effective in reducing the coupling between vertical and horizontal
betatron oscillations and is therefore quite useful to stabilize the
machine operation.

Some of the virtues of the NB option are quite evident: There is no
need to have short bunches (the condition $\sigma_z < \beta^*_y$ vital
for the HC option to avoid the hour glass effect does not apply here),
and the beam-beam parameter $\xi_y$ can be relaxed substantially (it
should be noted that the large $\xi_y$ parameter required for the HC
option has never been reached so far). Also
the synchrotron radiation is less of an issue due to the smaller beam
currents. This is certainly good news for the detectors close to the
beampipe (e.g. the silicon detectors). On the other hand, a new source
of background may arise from the Touschek effect~\cite{touschek}, a
kind of intra-beam Coulomb scattering, which couples betatron and
synchrotron oscillations. Due to this effect particles may be sent to
off-momentum orbits, causing particle loss and a severe decrease of the
beam lifetime. To counteract the shorter lifetimes (order of a few
minutes) an elaborate scheme is envisaged for SuperKEKB which keeps
the bunch currents almost constant by permanent re-injection at a
rate of about 50 Hertz. With the NB option another important physics
parameter may be chosen more favorably, i.e. the radius of the
beampipe, which may be as small as roughly 1 cm, whereas in the HC
option 1.5 cm had to be chosen. It is expected that the final design
for the SuperKEKB machine in its nano-beam version will be completed
in the fall of 2009.

\section{Detector Upgrade: BELLE-II}

Parallel to the machine studies, a Letter of Intent~\cite{belleloi}
for the Belle upgrade 
was issued in the year 2004, with a recent update from
2008~\cite{sbelle}.  The main (re)design goals are to cope with the
much higher physics rates and the much larger backgrounds to be
expected, as well as improving the overall physics performance.

\begin{figure}
  \includegraphics[height=0.47\textheight]{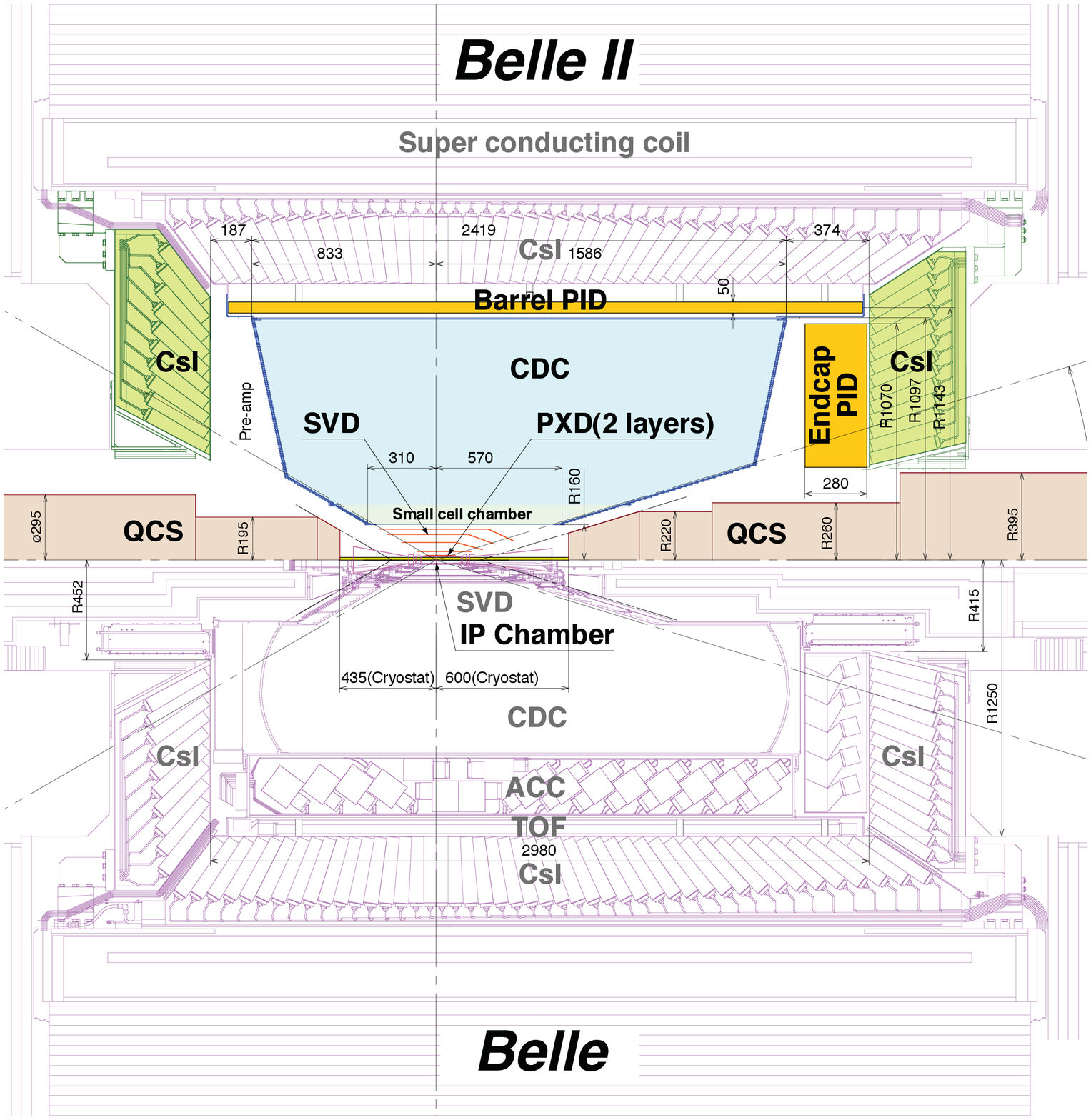}
  \caption{Upgrade scenario from the Belle detector (lower half) to the
  Belle-II detector (upper half).}
\label{belleii} 
\end{figure}

A comparison of Belle and its upgraded version (``Belle-II'') is shown
in a sideview in fig.~\ref{belleii}. We will first give a short
overview of the different detector components and then pick out two
specific systems which will be very important for the precise
measurement of the CP quantities. As a general side condition, the
performance of the new Belle-II detector should be as good or better
than Belle. This is a non-trivial requirement in view of the
anticipated very high background at the SuperKEKB machine, which is
estimated to be roughly an order of magnitude larger as compared to
the present KEKB machine (for details, see~\cite{sbelle}).

The tracking system in the ``old'' Belle detector consisted of 4 layer of
Si strip vertex detectors (SVD), followed by a Central Drift Chamber
(CDC). Due to the largely increased background, strip detectors are no
longer an option for the innermost Si layers. Instead a two-layer
pixel detector (PXD) for the innermost Si layers is planned (see
below). The SVD will be replaced entirely as well as the CDC: due to
the harsh backgrounds the inner radius of the CDC has to be moved out
and the two outer layers of the new SVD will cover the gap. The
momentum resolution of charged particles will be improved by extending
the CDC to a larger radius. Since the magnet and the barrel part of
the electromagnetic calorimeter (ECL) will not be changed, the
particle identification system has to be replaced by a thinner,
low material budget detector (see below). In the endcap parts of the ECL
the present CsI(Tl) crystals will be replaced by pure CsI that
provide faster signals, and the forward part of the KLM ($\kl$ and
muon detector) in the iron flux return yoke will be instrumented by
scintillator strips with SiPM readout, replacing the present RPCs.

To extend the physics reach for Belle-II, the $K/\pi$ separation
ability will be improved by a new particle identification (PID)
system. Two types of detectors are being proposed, one is a
time-of-propagation (TOP) counter for the barrel region (Barrel PID),
and the other a proximity-focussing Cherenkov ring imaging counter
with aerogel radiators (ARICH) for the endcap region (Endcap PID).     
\begin{figure}
  \includegraphics[width=\textwidth]{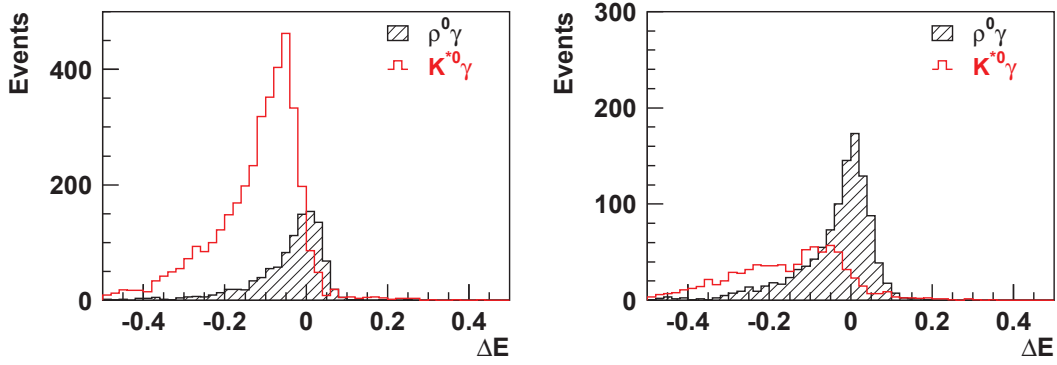}
  \caption{Comparison of the PID system performance of Belle (left) and
  Belle-II (right) for the decay $B^0 \ra \rho^0 \gamma$. The abscissa is
  $\Delta E= \Eb-\EB$ (in GeV) in the 
  $\Upsilon(4S)$ center-of-mass system (see text).} 
\label{fig:pid}
\end{figure}
The present time-of-flight and aerogel Cherenkov counters in the
barrel region of Belle are replaced by a TOP counter made from
quartz radiator bars (thickness 2 cm), in which the time of
propagation of Cherenkov photons is measured, which are internally
reflected and focussed onto micro-channel plate 
(MCP) PMTs at the end surfaces of the quartz bars. The MCP-PMTs have
an excellent time resolution of about 50 ps, so that the difference in
arrival times of the Cherenkov photons, radiated by  pions or
kaons, can be determined. This time difference results from the
different Cherenkov emission angles (for equal particle momenta) and
consequently different path lengths for the photons which undergo multiple 
internal reflections on their way to the ends of the quartz bars. 

The ARICH counters are located in front of the endcap ECL, where the
space is quite limited. For this reason a proximity focussing scheme
is envisaged for the Cherenkov photons, with an expansion thickness of
only 20 cm. In the present design three layers of silica aerogel, each
10 mm thick, are foreseen, with refractive indices varying
between 1.045 and 1.055, so that the photons emitted from the three
regions produce overlapping images on the photon detector surface. Since 
the photon detectors have to work in a strong magnetic field of 1.5
Tesla, candidates are a hybrid avalanche photon detector (HAPD) or a
MCP-PMT. 

An example of the expected performance of the new PID system relative
to the present Belle ACC system is shown in
fig.~\ref{fig:pid}. Here, the decay $B^0 \ra \rho^0 \gamma$ is studied at
the $\Upsilon(4S)$ resonance and the distribution of the quantity
$\Delta E=\Eb-\EB$ is chosen for reference, where $\Eb$ is
the beam energy in the $\Upsilon(4S)$ center-of-mass system and $\EB$ is
the energy of the $B$ meson, reconstructed from the two pions forming
the $\rho$ and the photon.  An overwhelming
background from $B \ra K^* \gamma$ is expected due to the CKM 
couplings (about a factor 40). While for the case of Belle (left side)
the background from misidentified kaons is very large, the improved PID in
Belle-II (right) largely reduces the background and shows a much clearer
signal.       
\begin{figure}
  \includegraphics[width=0.27\textwidth]{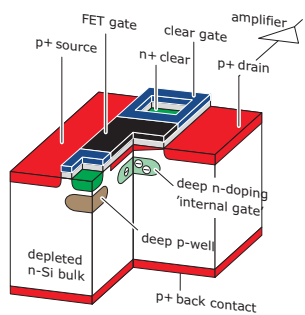}
  \caption{Principle of operation of the DEPFET pixel sensor.}
  \label{fig:depfet} 
\end{figure}

A second example for the improved instrumentation at Belle-II is the
new two-layer Si pixel detector, located closest to the beampipe, for
precise vertexing. As mentioned above, the expected strong increase of
background relative to the present KEKB machine, proportional to
$1/r^2$ where $r$ is the radial distance from the 
interaction region, creates large occupancies within the strip
detectors, making an efficient reconstruction of tracks and vertices
impossible. The solution is to use a pixel detector which
intrinsically provides three-dimensional space points. However,
pixel segmentation is not the only requirement to be fulfilled at
Belle-II. Due 
to the small (transverse) momenta of the $B$ decay products, the
momentum and vertex resolutions are dominated by multiple scattering,
so a pixel sensor should have a very low material budget. Furthermore, the
harsh radiation environment at the SuperKEKB factory requires
low-noise and radiation-hard technology, and the limited space between
the beampipe and the strip detector requires a low power
consumer. Finally, a pixel detector should be operational from the
start of the Belle-II running, which is envisaged for the year 2013. 

It turns out that there is only one mature technology at present
fulfilling all these requirements: The DEPFET pixel
sensor~\cite{depfet}, invented and developed at the
Max-Planck-Institute for Physics (MPI) in Munich, Germany. The name
DEPFET stands for $DEP$leted $F$ield $E$ffect $T$ransistor and combines
detection and amplification. The DEPFET principle of operation is
shown in fig.~\ref{fig:depfet}. A MOS field effect transistor is
integrated onto a fully depleted silicon substrate forming the
detector. By 
means of an additional $n$-implant underneath the transistor channel a
potential minimum for electrons is created. This potential well can be
considered as an internal gate of the transistor. A particle entering
the detector creates electron-hole pairs in the fully depleted silicon
substrate. The electrons (``signal'') are collected and stored in the
internal gate. The electron charges change the potential of the
internal gate, resulting in a modulation of the transistor channel
current. After readout of the channel current, the signal electrons
are removed by a positive voltage at the clear contact. Another read
cycle then establishes the baseline which is subtracted from the
current readout before the clear. The whole readout cycle
(read-clear-read) takes about 80 ns. Because
of the small capacitance of the internal gate, the noise in the DEPFET
is very low, about 100-200 electrons are expected when operating at
Belle-II. 

For a real detector a matrix of DEPFET pixels must be
constructed. Such matrices, using the SOI technique, where the sensor
wafer is bonded to a ``handling'' wafer, have already been
built and subjected to extensive beam tests. Matrices up to 512
$\times$ 512 pixels have been obtained, with pixel sizes of about 17
$\times$ 13 $\mu$m$^2$. For the DEPFET pixel detector (``PXD'') at
Belle-II typical pixel sizes would be around 50 $\times$ 50 $\mu$m$^2$.
An additional asset of the DEPFET technology is the fact that the
detector can be thinned down to about 50 $\mu$m thickness. The thinning
procedure of the handling wafer has already been demonstrated at the
MPI semiconductor laboratory.  With such a thin sensor still a signal
to noise ratio of about 40:1 can be reached. Finally, the radiation
hardness of the DEPFET matrix has been tested and found to be
satisfactory for a few years of running at SuperKEKB with full
luminosity. Further details of the DEPFET project at Belle-II can be
found elsewhere~\cite{depfetpage}.

\section{Conclusions}

Complementary to the LHC, a new generation of Super Flavor Factories
is being planned, probing the Standard Model at energy scales
beyond tens of TeV. Flavor physics continues to be the key to the
puzzle of the observed matter-antimatter asymmetry in the universe,
intimately connected to CP violation. The SuperKEKB project, together
with a substantially upgraded detector, Belle-II, will contribute to
this fascinating chapter of particle physics, taking first data  in
the year 2013, with the prospect of accumulating 50 times the presently
available data by the year 2020.


\begin{theacknowledgments}

The author would like to acknowledge the extremely stimulating and
pleasant atmosphere created by the organizers, especially Marvin
Marshak, of the CIPANP 2009 Conference. He also wishes to thank Tom
Browder, Yoshihiro Funakoshi, Katsunobu Oide, and Yoshi Sakai for
their careful reading of the manuscript and their valuable advice.

\end{theacknowledgments}



\bibliographystyle{aipproc}   


\IfFileExists{\jobname.bbl}{}
 {\typeout{}
  \typeout{******************************************}
  \typeout{** Please run "bibtex \jobname" to optain}
  \typeout{** the bibliography and then re-run LaTeX}
  \typeout{** twice to fix the references!}
  \typeout{******************************************}
  \typeout{}
 }


\end{document}